\documentclass[pra,twocolumn,superscriptaddress,amsmath,amssymb,floatfix,noshowpacs,nofootinbib]{revtex4-2} 
\usepackage{graphicx}   
\usepackage{dcolumn}    
\usepackage{bm}         

\begin{document}
\title{\textbf{Complementary mechanisms of critical quantum metrology}}
\author{Nikita Leppenen}
\affiliation{Department of Chemical \& Biological Physics, Weizmann Institute of Science, Rehovot 7610001, Israel}
\author{Shaked Levy}
\affiliation{Department of Chemical \& Biological Physics, Weizmann Institute of Science, Rehovot 7610001, Israel}
\author{Ephraim Shahmoon}
\affiliation{Department of Chemical \& Biological Physics, Weizmann Institute of Science, Rehovot 7610001, Israel}
\date{\today}

\begin{abstract}
Critical metrology conventionally exploits enhanced susceptibility to a system parameter near a phase transition. Here we identify a complementary mechanism for critically enhanced estimation of intrinsic system parameters via suppressed quantum noise rather than enhanced response. We demonstrate both mechanisms in coherently radiating spin states (CRSS)—correlated states of $N$ spin-$1/2$ particles defined as asymptotic eigenstates of the collective-spin lowering operator with complex eigenvalue $r e^{-i\varphi}$. At the critical value $r=1$, CRSS exhibit a second-order transition. We show that near criticality, estimation of $r$ benefits from enhanced susceptibility despite increased noise, whereas estimation of $\varphi$ benefits from suppressed quantum noise without enhanced response. Remarkably, both yield the same estimation error $\sim N^{-2/3}$, improving over the standard quantum limit $N^{-1/2}$. For $\varphi$, this establishes quantum-enhanced metrology of phases intrinsic to the critical system, rather than phases externally encoded by interferometry. We demonstrate these results for CRSS realizations in both dissipative (superradiant) and Hamiltonian (Ising) systems relevant to cavity QED and trapped ions, where fast preparation of CRSS is shown possible thanks to gap opening. Applications to metrology of coupling rates and phases or optical lengths are discussed.
\end{abstract}

\maketitle

The central goal of quantum metrology is the estimation of a physical parameter $x$ from measurements of a system observable $\hat{A}$ \cite{QsenseREV,QmetrologyREV,OberthalerREV}. The sensitivity, or estimation error, is given by the error propagation formula, $\delta x=\delta A/ |\partial_x \langle \hat{A}\rangle|$, which can be viewed as an inverse signal-to-noise ratio, with the signal given by the susceptibility of $\hat{A}$ to changes in $x$ and the noise by $\delta A=(\mathrm{Var}[\hat{A}])^{1/2}$.
 Enhanced sensitivity (small $\delta x$) can therefore be achieved either by an enhanced signal $|\partial_x \langle \hat{A}\rangle|$ or by suppressed noise $\delta A$ \cite{YEr}.

Critical metrology exploits the enhanced sensitivity of a system near a critical point, e.g., in a dissipative or quantum phase transition
\cite{ParisSalvatori2014,zanardi2008quantum,Garbe2020,Porras2017,IvanovPorras2023,Montenegro2023,chu2021dynamic,Marzolino2014,Heugel2019,ZubinQCM1,LesanovskyCQM2016,LesanovskyCQM2024,Plenio2022,Adams2022,Weimer2018,Clerk2022,ClerkBelliardo2026,Hotter2024,ivanov2020,DiFresco2022,MontenegroREV}.
Conventionally, this enhancement relies on the large susceptibility of a system observable --- often an order parameter (e.g. magnetization)--- to variations of a parameter $x$ (e.g. field strength) near its critical value \cite{GietkaREV}. The resulting metrological gain is therefore attributed to the critically enhanced signal, $|\partial_x \langle \hat{A}\rangle|$, even when critical fluctuations increase the measurement noise. Such enhancement mechanism can be often understood already at the semiclassical, mean-field level \cite{Gammelmark2011,yoav}.

Here we put forward a complementary mechanism of critical metrology that relies instead on suppression of the noise $\delta A$, without enhanced susceptibility. In this case, enhanced sensitivity $\delta x$ over the standard quantum limit (SQL) $N^{-1/2}$ originates directly from quantum correlations at criticality that suppress the fluctuations $\delta A$ of the measured observable.
Correlated critical states exhibiting suppressed quantum noise were studied in a variety of systems \cite{Frerot2018,Fazio2002,Alejandro,Vidal2004LMG}, and were considered as input states for interferometric metrology: namely, the estimation of an \emph{external} parameter encoded on the input state by a subsequent interferometric transformation \cite{MontenegroREV,GietkaREV}.
Here, instead, we show that critically suppressed quantum noise can enhance metrology of parameters \emph{intrinsic} to the critical system itself, offering distinct possibilities in sensing its internal properties or control fields.

We demonstrate these concepts in critical systems of $N$ spins described by coherently radiating spin states (CRSS). CRSS are defined as the asymptotic eigenstates of the collective-spin lowering operator characterized by a complex eigenvalue $re^{-i\varphi}$: they exhibit a second-order transition at $r=1$ and spin-squeezing entanglement \cite{CRSS}. First, we show analytically that the two complementary mechanisms of critical metrology coexist in CRSS: estimation of $r$ benefits from enhanced susceptibility despite increased noise, whereas estimation of $\varphi$ benefits from suppressed (squeezed) quantum noise without enhanced response. Remarkably, both yield the same beyond-SQL scaling $\delta r,\delta\varphi\sim N^{-2/3}$. Second, we demonstrate this CRSS metrological framework in dissipative (superradiant) and Hamiltonian (Ising) systems relevant to cavity QED and trapped ions, enabling quantum-enhanced sensing of intrinsic phases, coupling rates, and optical lengths without subsequent interferometric encoding. Finally, we show that fast preparation of CRSS is possible thanks to gap opening near criticality, avoiding the usual critical slowing down.

\begin{figure}[t]
  \centering
  \includegraphics[width=\columnwidth]{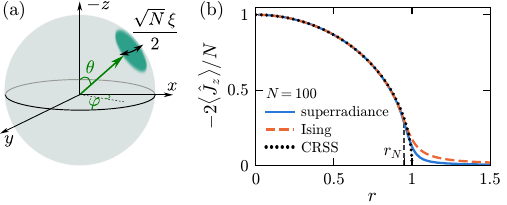}
  \caption{(a) CRSS represented on the collective-spin Bloch sphere ($-z$ denoting unexcited spins). The CRSS eigenvalue $re^{-i\varphi}\equiv \sin\theta e^{-i\varphi}$ determines the angles $(\theta,\varphi)$ of the mean-spin vector of length $j=N/2$. The ellipse-shaped contour of the variance around the mean-spin showcases the CRSS spin squeezing $\xi^2=\sqrt{1-r^2}=\cos\theta$. (b) Second order phase transition of CRSS seen in the magnetization $\langle \hat{J}_z \rangle$. The critical point $r\rightarrow 1$ offers opportunities for metrology. CRSS predictions are seen to agree asymptotically with results obtained by stable states of the realizations from Fig. 2 [Eqs. (\ref{ME}), (\ref{H})].
  }\label{fig1}
\end{figure}

\emph{CRSS and physical realizations.---}
We begin with a brief account of CRSS in collective spin systems. Consider $N \gg 1$ spins with spin-1/2 lowering operators $\hat{\sigma}^-_n$ and population-inversion operators $\hat{\sigma}^z_n$ for spins $n=1,...,N$. Within the maximal-spin representation, $j=N/2$, the CRSS are defined as the asymptotic eigenstates of the collective-spin lowering operator $\hat{J}_-=\sum_n \hat{\sigma}^-_n$ \cite{CRSS},
\begin{eqnarray}
\hat{J}_- |r,\varphi\rangle = j r e^{-i\varphi} |r,\varphi\rangle,
\quad
|r,\varphi\rangle=\sum_{m=-j}^{j} a_m  |j,m\rangle,
\label{CRSS}
\end{eqnarray}
with the complex eigenvalue $r e^{-i\varphi}$ (normalized to $j=N/2$) and the expansion coefficients $a_m(r,\varphi)$ in the eigenstates $|j,m\rangle$ of the magnetization operator $\hat{J}_z=(1/2)\sum_n \hat{\sigma}_n^z$ (with eigenvalues $m\in\{-j,j\}$). While mathematically, such exact states do not exist for finite $N$ (apart from $r=0$), they can be constructed asymptotically for $N\rightarrow \infty$, and practically for any large $j=N/2$, as (see \cite{CRSS} for more complete definitions),
\begin{eqnarray}
&&a_m = e^{-i(m+j)\varphi}\sqrt{P_m(r)}, \quad P_m \approx \frac{1}{\sqrt{2\pi w^2}} e^{-(m-m_0)^2/2w^2},
\nonumber\\
&& w^2(r)=\frac{j r^2}{2\sqrt{1-r^2}}, \quad m_0(r)=-j\sqrt{1-r^2}.
\label{am}
\end{eqnarray}
For a finite $N\gg 1$, this construction constitutes an eigenstate of $\hat{J}_-$ for any eigenvalue within the validity region $0\leq r< r_N \sim \sqrt{1-\left(\frac{3}{N}\right)^{2/3}}\rightarrow 1$, up to an exponentially small error in $j=N/2$.

Notably, since $|\langle \hat{J}_- \rangle|=(N/2) r\equiv (N/2)\sin\theta$, the upper limit $r_N \rightarrow 1$ corresponds to the equator of the collective-spin Bloch sphere ($\theta \rightarrow \pi/2$, Fig. 1a): thus, CRSS describe collective physics of a macroscopic excitation fraction (up to half filling), well beyond the linearized Holstein-Primakoff approximation. This is demonstrated by two key properties of CRSS: (i) The magnetization $\langle \hat{J}_z\rangle\approx m_0=-(N/2)\sqrt{1-r^2}$ exhibits a second-order phase transition at $r=1$ (softened at $r\sim r_N$ for finite $N$, Fig. 1b), with diverging susceptibility. (ii) For $r>0$, quantum correlations suppress fluctuations perpendicular to the mean-spin direction, as quantified by the spin-squeezing parameter $\xi^2=\sqrt{1-r^2}$ ($\xi^2<1$ signifying entanglement, Fig. 1a), with maximal suppression near criticality $r=r_N\rightarrow 1$.

These two key properties already underlie the two complementary mechanisms for critical metrology of the CRSS parameters $r$ and $\varphi$. Before we analyze this in detail, we briefly discuss physical realizations of CRSS as stable states. First, consider the dissipative system of laser-driven two-level atoms that radiate collectively through their common coupling to a fast-decaying cavity mode (Fig. 2a). The corresponding master equation for the many-atom density matrix $\hat{\rho}$ is given by \cite{CAR,HL},
\begin{eqnarray}
&&\hspace{-0.4em}\frac{d\hat{\rho}}{d t}=-\frac{i}{\hbar}\left[\hat{H}, \hat{\rho}\right]
+\gamma\hat{J}_- \hat{\rho} \hat{J}_+ - \frac{\gamma}{2}\left(\hat{J}_+\hat{J}_-\hat{\rho}+\hat{\rho}\hat{J}_+\hat{J}_-\right),
\label{ME}\\
&&\hspace{-0.4em}\hat{H}/\hbar=\Delta\hat{J}_+\hat{J}_- + \left(\Omega \hat{J}_+ + \Omega^{\ast} \hat{J}_-\right),
\;
\Omega_c=\frac{N}{4}\sqrt{\gamma^2+4\Delta^2}.
\nonumber
\end{eqnarray}

Here $\gamma$ and $\Delta$ are cavity-mediated collective decay and dipole-dipole interaction, and $\Omega=|\Omega|e^{i\phi}$ is the drive amplitude.
It can be shown that the CRSS with  $r\ = |\Omega|/\Omega_c$ and $\varphi=\arctan(2\Delta/\gamma)-\phi-\pi/2$ forms the steady state of Eq. (\ref{ME}) for $r<1$ \cite{CRSS} and underlies the dissipative Dicke phase transition at $r=1$ (Fig. 1b) \cite{CAR,HL,TOMpt,DRUMMOND1978160,DRUMMOND1980,LAW,LAR,BAR,Lee2014,KES,RABwigner}.

\begin{figure}[!tb]
  \centering
  \includegraphics[width=\columnwidth]{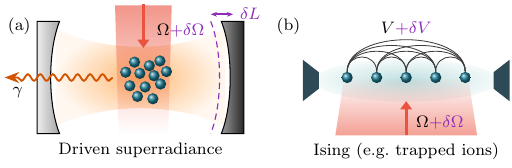}
  \caption{CRSS in collective spin systems. (a) Cavity superradiance realization: collective radiation of atoms (``spins”) via the common cavity mode leads to their dissipative coupling as described by the driven-dissipative Dicke model from Eq. (\ref{ME}). The steady state is a CRSS. (b) Trapped-ion Ising realization: internal states of the ions (``spins") are coupled to a common collective motional mode that mediates all-to-all interactions. This realizes the globally interacting Ising Hamiltonian, Eq. (\ref{H}), whose ground state approaches a CRSS.
  }\label{fig2}
\end{figure}

Turning to interacting Hamiltonian systems, consider the Ising coupling at strength $V$ identical for all interacting spin pairs, as described by $-V\hat{J}_z^2$. In trapped ions, such all-to-all coupling between the ions' pseudo-spins is mediated by common center-of-mass motion \cite{MONrev,BlattRoos,Bollinger,Ozeri} (Fig. 2b), whereas in neutral atoms this may be achieved via a Rydberg-dressing potential \cite{POHrd,Schleier_rb_2023,Zeiher_rd_2025,Kaufman_rd_2025}. Together with transverse and longitudinal fields $V (N/2) r e^{-i\varphi}$ and $V \beta$, respectively, we obtain the Hamiltonian
\begin{eqnarray}
\hat{H}=V\left[-\hat{J}_z^2+\beta\hat{J}_z +(N/2) r (e^{-i\varphi}\hat{J}_-+e^{i\varphi}\hat{J}_+)\right].
\label{H}
\end{eqnarray}
For transverse field $r<1$ (and $\beta \ll N$) the ground state approaches the CRSS $|r,\varphi\rangle$ \cite{CRSS}, which underlies the quantum phase transition of this global Ising (Lipkin-Meshkov-Glick) model at $r=1$ (Fig. 1b) \cite{Heiss2005LMGsemi,Vidal2005LMG,Liberti2010LMG}.

In both realizations, Eqs. (\ref{ME})--(\ref{H}), the CRSS parameters $r,\varphi$ directly relate to system parameters like fields and couplings, which may themselves depend on other physical quantities of interest. Critical metrology of $r,\varphi$ thus enables enhanced estimation of these quantities.

\emph{Theory of critical metrology with CRSS.---}
We now turn to the general analysis of critical metrology of CRSS, which is subsequently applied to the physical realizations above.
First, we note that the mean magnetization under CRSS, $\langle \hat{J}_z \rangle\approx -(N/2)\sqrt{1-r^2}$, becomes highly sensitive to changes in $r$ when approaching the critical point $r=1$. Inserting the resulting susceptibility $\partial_r \langle \hat{J}_z \rangle$ and the measurement uncertainty $\text{Var}[\hat{J}_z]=w^2$ [Eq. (\ref{am})] into $\delta r =(\text{Var}[\hat{J}_z])^{1/2}/|\partial_r \langle \hat{J}_z \rangle|$, we obtain the sensitivity
\begin{eqnarray}
\delta r = \frac{\left(1-r^2\right)^{1/4}}{\sqrt{N}},
\quad
\delta r_N = \left.\delta r\right|_{r=r_N} \sim N^{-2/3}.
\label{dr}
\end{eqnarray}
The first expression exhibits the expected sensitivity enhancement, with an error reduction $(1-r^2)^{1/4}$ beyond the SQL, favoring to work as close to $r=1$ as possible. Recall, however, that CRSS theory breaks down beyond $r=r_N$, where the transition is softened for finite $N$, setting $r=r_N\sim \sqrt{1-\left(\frac{3}{N}\right)^{2/3}}$ as the optimal working point. This yields the CRSS prediction for the optimal sensitivity $\delta r_N\sim N^{-2/3}$, surpassing the SQL $N^{-1/2}$.

Importantly, while our choice of the magnetization estimator for $r$ was based on the intuitive argument of large susceptibility, we note that:
(i) this estimator is effectively optimal, in the sense that its error approaches the quantum Cram\'{e}r-Rao bound, $\delta r \sim 1/\sqrt{F_r}$ (with $F_r$ the quantum Fisher information for estimating $r$) ---
either saturating it asymptotically (for fixed $r$) or reaching it within an $O(1)$ factor (for $r=r_N$); we verify this by calculating $F_r$ directly (End Matter).
(ii) The measurement noise of this estimator in CRSS, $\text{Var}[\hat{J}_z]=w^2$, scales as $N^{4/3}$ near criticality (at $r=r_N$), which is worse than the $O(N)$ noise-scaling of an uncorrelated state. In Fig. 1a, this is seen by anti-squeezed fluctuations along $z$.

\begin{table}[t]
\label{tab:mechanisms}
\begin{ruledtabular}
\begin{tabular}{lcc}
Quantity
& $r$ estimation
& $\varphi$ estimation \\
\hline
Parameter $x$
& $r$ & $\varphi$ \\
Observable $\hat A$
& $\hat J_z$ & $\hat J_y$ \\
Signal $|\partial_x\langle\hat A\rangle|$
& $N^{4/3}$ & $N$ \\
Noise $\delta A=\sqrt{\mathrm{Var}(\hat A)}$
& $N^{2/3}$ & $N^{1/3}$ \\
Error $\delta x =\delta A/|\partial_x\langle\hat A\rangle|$
& $N^{-2/3}$ & $N^{-2/3}$ \\
\hline
\shortstack{Origin of\\enhancement}
& \shortstack{Enhanced\\susceptibility}
& \shortstack{Suppressed\\quantum noise} \\
\end{tabular}
\caption{Complementary mechanisms of critical metrology in CRSS. Finite-size scaling at criticality, $r=r_N$, where $1~-~r_N^2\sim N^{-2/3}$. Considering the standard reference case of independent spins, with the scalings $N$ and $\sqrt{N}$ for signal and noise, respectively, we observe that while the estimation errors for both $r$ and  $\varphi$ exhibit the same overall enhancement, the first does so through enhanced signal (despite increased noise), whereas the second via suppressed quantum noise. Full results appear in text and Eqs.~(\ref{dr}),~(\ref{dphi}).}
\end{ruledtabular}
\end{table}

Next, we consider the estimation of the CRSS phase parameter $\varphi$.
Although $\varphi$ is intrinsic to the critical state rather than accumulated during interferometric interrogation, its variation is mathematically equivalent to a rotation generated by the operator $\hat{G}=\hat J_z+j$, since $|r,\varphi + d\varphi\rangle= e^{-i \hat{G}d\varphi} |r,\varphi\rangle$ [see Eq. (\ref{am})]. This suggests that small changes in $\varphi$ are best detected by measuring the orthogonal quadrature operator $\hat{J}_{\varphi}=\cos\varphi \hat{J}_y-\sin\varphi\hat{J}_x$, as clearly seen at criticality ($r\rightarrow 1$) where the mean spin approaches the equator of the Bloch sphere $\theta \rightarrow \pi/2$ (Fig. 1a). Choosing $\varphi=0$ as our working point, we therefore estimate $\varphi$ through measurements of $\hat{J}_y$.
This yields the estimation error $\delta \varphi=(\text{Var}[\hat{J}_y])^{1/2}/|\partial_{\varphi} \langle \hat{J}_y \rangle|$, finding
\begin{eqnarray}
\delta \varphi = \frac{(1-r^2)^{1/4}}{\sqrt{N}r},
\quad
\delta \varphi_N = \left.\delta \varphi\right|_{r=r_N} \sim N^{-2/3}.
\label{dphi}
\end{eqnarray}
Here we again verified that the intuitive choice of $\hat{J}_y$ yields an optimal estimator that asymptotically saturates the quantum Cram\'{e}r-Rao bound $\delta \varphi=1/\sqrt{F_{\varphi}}$ for estimation of $\varphi$ (End Matter).
The sensitivity enhancement increases with $r$, with the optimal scaling deduced at criticality $r=r_N$ as $\delta \varphi_N \sim N^{-2/3}$.

Since the signal $|\partial_{\varphi} \langle \hat{J}_y \rangle|= (N/2) r$ is not enhanced at criticality, the gain in sensitivity is originated in reduced critical fluctuations $\text{Var}[\hat{J}_y]=(N/4)\xi^2$ related to the CRSS spin squeezing  $\xi^2=\sqrt{1-r^2}$. Intuitively, this is understood by noting that, as the system approaches criticality $r \rightarrow 1$, $\hat{J}_y$ increasingly overlaps with the squeezed quadrature of CRSS  (Fig. 1a).
Therefore, here quantum squeezing enhances estimation of an intrinsic parameter $\varphi$ of the critical state, rather than an externally imprinted interferometric phase, opening distinct possibilities for sensing physical parameters encoded in the system itself.

To conclude, while the sensitivities of both CRSS parameters scale the same at criticality, as $N^{-2/3}$, it is important to distinguish the origins of their metrological gain over the SQL: this distinction is not revealed by the sensitivity (or Fisher information) alone, but rather by the separate examination of its signal and noise components, as summarized in Table I.
The complementarity of distinct enhancement mechanisms is also reflected in the corresponding measured observables, $\hat{J}_z$ and $\hat{J}_y$, which constitute, respectively, the anti-squeezed and squeezed quadratures of CRSS near criticality (Fig. 1a).

\begin{figure}[!tb]
  \centering
  \includegraphics[width=\columnwidth]{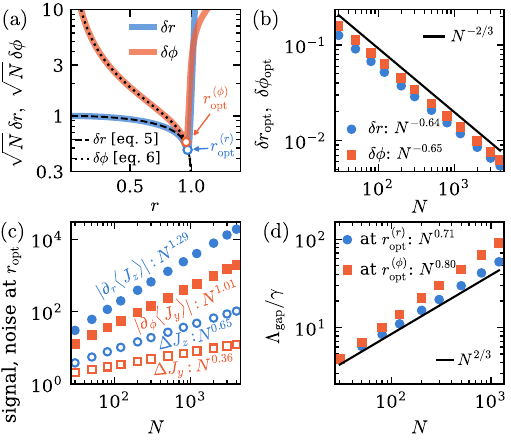}
  \caption{Superradiance realization: exact steady state of Eq. (\ref{ME}). (a) Estimation errors $\delta r, \delta \varphi$ reach their optimal (minimal) values at respective field strengths $r_{\text{opt}}^{(r,\varphi)}$ that approach the CRSS prediction $r_N$ (here $N=400$). (b) Scaling of the optimal values of $\delta r, \delta \varphi$ with $N$ is in excellent agreement with the CRSS prediction $N^{-2/3}$ for both $r$ and $\varphi$. (c) Complementary mechanisms of critical metrology revealed by the distinct scalings of signal and noise with $N$, which agree well with those of Table I.
  (d) Liouvillian gap $\Lambda_{\text{gap}}$ of Eq. (\ref{ME}) as a function of $N$ (evaluated at $r_{\text{opt}}^{(r,\varphi)}$). The exhibited scalings are consistent with the estimation $N^{2/3}$ of the relaxation rate (considering finite $N$), implying that steady-state preparation of the critical state is not limited by critical slowing down.
  }\label{fig3}
\end{figure}

\emph{Realization in superradiant dissipative phase transition.---}
We now show how the above theory applies for critical metrology in the superradiance case of Fig. 2a. Recall that the steady state of Eq. (\ref{ME}) is predicted to be well approximated by the CRSS with eigenvalue amplitude $|\Omega|/\Omega_c$ and phase $\arctan (2\Delta/\gamma)-\phi-\pi/2$: fixing the drive phase $\phi=-\pi/2$ and working at $\Delta/\gamma\ll 1$ (arranged by controlling the atom-cavity detuning, see below), we then study critical metrology of the normalized drive strength $r\equiv|\Omega|/\Omega_c$ and dipole-dipole interaction $\varphi\equiv 2\Delta/\gamma$ and compare them to the predictions for the related CRSS parameters.

To this end, we numerically solve Eq. (\ref{ME}) for the steady-state density matrix $\hat{\rho}_s$ at different drive strengths $r$ around $\varphi=0$, and evaluate the sensitivities $\delta r$ and $\delta\varphi$ from error propagation for measurements of $\hat{J}_z$ and $\hat{J}_y$, respectively (End Matter). In Fig. 3a, we plot these sensitivities versus $r$ for $N=400$, observing their decrease toward optimal (minimal) values $\delta r_{\text{opt}}(N)$ and $\delta\varphi_{\text{opt}}(N)$ near criticality, $r\sim r_N$, in agreement with the CRSS predictions of Eqs. (\ref{dr}) and (\ref{dphi}). Repeating this for different $N$, we obtain the beyond-SQL scaling $\delta r_{\text{opt}},\delta\varphi_{\text{opt}}\sim N^{-2/3}$ (Fig. 3b), consistent with the CRSS predictions $\delta r_N$ and $\delta\varphi_N$. The result for $r$ is also consistent with previous numerical studies of (\ref{ME}) \cite{IvanovPorras2023,Montenegro2023}.

To demonstrate the emergence of the two complementary mechanisms of critical metrology in the cavity realization, we plot in Fig. 3c the signal and noise components for both $r$ and $\varphi$, obtained with $\hat{\rho}_s$ at criticality. Their scalings with $N$ are seen to exhibit excellent agreement with the CRSS predictions of Table I.


The estimation of $r$ and $\varphi$ requires measurements of $\hat{J}_z$ and $\hat{J}_y$, respectively. In this realization, $\hat{J}_z$ is the atomic population inversion, measurable via off-axis fluorescence \cite{Ferioli2023,yoav} or shelving \cite{TOMpt}. To access $\hat{J}_y$, one may either map it onto $\hat{J}_z$ via a rapid $\pi/2$ pulse, or transfer the excited state to a stable state for subsequent measurement.

All these measurements should be performed after the system relaxed to its critical steady state.
We can deduce the relaxation rate $\Lambda$ from Eq. (\ref{ME}) using a mean-field analysis, finding $\Lambda/\gamma \sim N^{2/3}$ at criticality $r=r_N$ (End Matter). More formally, $\Lambda$ is bounded by the ``Liouvillian gap" $\Lambda_{\text{gap}}$, given by the first non-zero eigenvalue of the Liouvillian of Eq. (\ref{ME}). In Fig. 3d, we plot the numerically calculated $\Lambda_{\text{gap}}$ at criticality, observing consistency with the critical scaling $N^{2/3}$ (up to finite-size effects).
This ``gap opening" with $N$ implies that the steady-state preparation time does not increase with $N$ at criticality, and is hence not limited by usual critical slowing down of relaxation \cite{MontenegroREV,GietkaREV,yoav}. The resulting metrological protocol may ultimately be limited by $N$-independent initialization processes unrelated to the collective physics considered here.
Notably, our result $\Lambda \sim N^{2/3}$ is in contrast to previous estimations of critical gap closing in this model \cite{Montenegro2023}. The difference is due to the $N$-independent atom-cavity interaction taken here, in accordance with actual realizations for which the cavity volume remains fixed when atoms are added \cite{TOMpt,TOMs1,TOMs2,rivero_quantum_2023,SKtw1,SKtw2,MAJ}.

\emph{Metrological applications: superradiance realization.---}
Considering the atom-cavity coupling $g$ (identical for all atoms) and detuning $\delta_c$, and the cavity damping rate $\kappa$ and laser drive $\Omega_L$, we obtain the relations $\Delta-i\gamma/2=g^2/(\delta_c+i\kappa/2)$ and \mbox{$\Omega=\Omega_L$ or $\Omega=-g\Omega_L/(\delta_c+i\kappa/2)$}, depending, respectively, on whether the laser enters perpendicular to the cavity axis (as in Fig. 2a) or in parallel through its mirrors \cite{HL}.
For the latter, this gives the normalized drive strength $r=(|\Omega_L|/g)(2/N)$, whose metrology translates to critical metrology of either the dipole coupling or the driving-field strength, with $|\delta \Omega_L|/|\Omega_L|, \delta g /g \sim \delta r \sim N^{-2/3}$.
Alternatively, for the perpendicular configuration, where the phase $\phi$ of $\Omega=\Omega_L$ is independently fixed by the input laser, the metrology of $\varphi=2\Delta/\gamma$ around $\Delta=0$ demonstrated above can translate to sensing of $\delta_c$: this entails critical metrology of the cavity resonance frequency, and in turn, of changes in cavity length $L$ (Fig. 2a). For the latter, we obtain the sensitivity (per wavelength $\lambda$), $\delta L /\lambda \sim \delta \varphi /\mathcal{F} \sim N^{-2/3}/\mathcal{F}$ (End Matter), exhibiting critical enhancement $N^{-2/3}$ in addition to the enhancement from cavity finesse $\mathcal{F}$.
This illustrates new possibilities offered by critical noise suppression based on sensing intrinsic system parameters instead of interferometry: here, critical spin squeezing enables quantum-enhanced sensing of optical length $L$ (and effects influencing it), without requiring an optical interferometer or squeezed light.
The same critical enhancement applies for metrology of the laser-drive phase $\phi=-\varphi+\text{const.}$ --- either in the parallel or perpendicular configuration (fixing $\delta_c$ in the latter) --- e.g. in relation to sensing the optical path traversed by the laser (length, refractive index).

\emph{Realization in Ising quantum phase transition.---}
We now consider critical metrology of the normalized transverse-field amplitude $r$ and phase $\varphi$ in the Hamiltonian Ising model, Eq. (\ref{H}), whose ground state is predicted to be a CRSS. In analogy to the analysis of the dissipative realization, we numerically obtain the ground state of $\hat{H}$ (fixing $\beta=1$), and evaluate the estimation errors from measurements of $\hat{J}_z$ and $\hat{J}_y$. The optimal sensitivities occur near $r=r_N$ and exhibit the predicted $N^{-2/3}$ scaling (Figs. 4a,b). Moreover, the signal and noise scalings (Fig. 4c) confirm the two complementary mechanisms of critical enhancement from Table I.

Considering ion trap realizations \cite{MONrev,BlattRoos,Bollinger,Ozeri}, where the transverse field $\Omega=|\Omega|e^{-i\varphi}$ may represent a laser amplitude in a Raman scheme, we obtain that $r=2\hbar|\Omega|/(V N)$ encodes the ratio between field strength and Ising coupling $V$. Then, enhanced critical metrology of $r$ translates to that of either field or coupling strength, with $\delta |\Omega|/|\Omega|, \delta V/V\sim \delta r\sim N^{-2/3}$ (Fig. 2b). Since $V$ is mediated by the center-of-mass motional mode, this may enable sensing of charges or the electric environment.

\begin{figure}[!tb]
  \centering
  \includegraphics[width=\columnwidth]{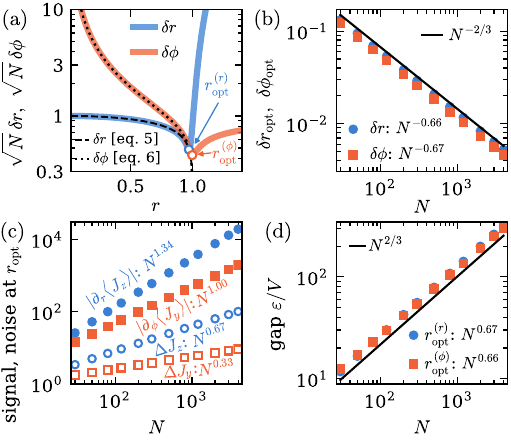}
  \caption{Ising realization: exact ground state of Hamiltonian (\ref{H}). (a) Estimation errors $\delta r, \delta \varphi$ reach their optimal (minimal) values at respective field strengths $r_{\text{opt}}^{(r,\varphi)}$ that approach the CRSS prediction $r_N$ (here $N=400$). (b) Scaling of the optimal values of $\delta r, \delta \varphi$ with $N$ is in excellent agreement with the CRSS prediction $N^{-2/3}$ for both $r$ and $\varphi$. (c) Complementary mechanisms of critical metrology revealed by the scalings of signal and noise with $N$, which are seen to agree well with those of Table I.
  (d) Energy gap $\varepsilon$ of Hamiltonian (\ref{H}) as a function of $N$ (evaluated at $r_{\text{opt}}^{(r,\varphi)}$). The exhibited scaling $\sim N^{2/3}$ suggests that adiabatic preparation of the critical state is not limited by critical slowing down (see also End Matter).
  }\label{fig4}
\end{figure}


As in the dissipative realization, the metrology protocol relies on preparing the critical state --- here the ground state of Hamiltonian (\ref{H}): this can be achieved by adiabatically varying $r$ from $r=0$, where the ground state is easily prepared, to $r\approx r_N$. A common bottleneck in critical systems is the closing of the energy gap $\varepsilon$ between the ground and first excited state, which limits the adiabatic preparation rate and leads to critical slowing down \cite{MontenegroREV,GietkaREV,Garbe2020,lu2026critical,Rams2018,Gietka2021}. Here, however, the gap opens with $N$ as $\varepsilon/V\sim N^{2/3}$ \cite{shaked}, as verified by exact diagonalization of Hamiltonian (\ref{H}) (Fig. 4d). This suggests that adiabatic preparation is not limited by critical slowing down, as further supported by an analysis of the adiabatic protocol (End Matter). Notably, this gap opening originates from collectivity and the $N$-independent pairwise interaction strength $V$, consistent with finite-size realizations such as trapped ions \cite{MONrev,BlattRoos,Bollinger,Ozeri}.

\emph{Discussion.---} We have demonstrated two complementary mechanisms of critical metrology in collective spin systems, revealing how critically suppressed quantum fluctuations can enhance the estimation of intrinsic system parameters without enhanced susceptibility. Beyond the sensing applications considered here, our results motivate the search for similar noise-driven metrological enhancement in other critical many-body systems. More broadly, they raise the question of the genuinely quantum origin of critical metrological enhancement, particularly in contrast to susceptibility-driven mechanisms that can sometimes be understood already at the semiclassical or mean-field level.

\begin{acknowledgments}
We acknowledge insightful discussions with James Thompson, and financial support from the Israel Science Foundation (ISF), the Minerva Stiftung with funding from the Federal German Ministry for Education and Research, the Shimon and Golde Picker Weizmann Annual Grant, the Center for New Scientists at the Weizmann Institute of Science, the Israeli Council for Higher Education - VATAT (QERNEL program), and the Dewey David Stone and Harry Levine Career Development Chair. This research is made possible in part by the historic generosity of the Harold Perlman Family.
%
\end{acknowledgments}

\makeatletter
\auto@bib@empty
\begingroup
\let\auto@bib@innerbib\@empty
\makeatother

\endgroup

\newpage

\section*{End Matter}
\emph{Quantum Fisher information calculations.---}
The quantum Fisher information (FI) of a pure state $|\psi(x)\rangle$ with respect to its internal parameter $x$ is given by \cite{MontenegroREV,GietkaREV} $F_x=4(\langle \partial_x \psi(x)|\partial_x \psi(x)\rangle - |\langle \partial_x \psi(x)|\psi(x)\rangle|^2)$. For a CRSS, since the state coefficients $a_m(r,\varphi)$ depend on $r$ only through their magnitudes $\sqrt{P_m(r)}$ [Eq. (\ref{am})], we generally find $F_{r}=\sum_m P_m(r)\left(\partial_r [\ln P_m(r)]\right)^2$, i.e. the quantum FI is equivalent to the classical FI of the distribution $P_m$. Using $P_m$ from  Eq. (\ref{am}) we obtain
\begin{eqnarray}
\hspace{-1.1em}F_r=\frac{(\partial_r m_0)^2+2(\partial_r w)^2}{w^2}=\frac{N}{\sqrt{1-r^2}}+\frac{(2-r^2)^2}{2r^2(1-r^2)^2}.
\label{Fr}
\end{eqnarray}
For $r<1$ independent of $N$, the first term is dominant ($N\gg 1$) and is identical to $(1/\delta r)^2$ from Eq. (\ref{dr}). For $r=r_N$ both terms scale as $N^{4/3}$ and are proportional to $(1/\delta r_N)^2$ from Eq. (\ref{dr}).
For $\varphi$, the dependence of the CRSS coefficients $a_m(r,\varphi)$ is only through the linear phase $e^{-i(m+j)\varphi}$, so that the quantum FI formula gives, $F_{\varphi}=4 \text{Var}[\hat{J}_z]=4w^2=Nr^2/\sqrt{1-r^2}$, identical to $(1/\delta \varphi)^2$ from Eq. (\ref{dphi}).

\emph{Sensitivity calculations for physical realizations.---}
Considering the superradiance and Ising realizations, we calculate numerically the stable state $\hat{\rho}(r,\varphi)$ --- given, respectively, by the steady state of Eq. (\ref{ME}) or the ground state of Hamiltonian (\ref{H}) --- for $r$ and $\varphi$ as defined for each model, and for a given $N$, using the basis states $|j=N/2,m\rangle$. To obtain the sensitivity for $r$ at a desired $r$ and $\varphi=0$, we calculate $\hat{\rho}(r,0)$ and $\hat{\rho}(r\pm dr,0)$ with a sufficiently small step $dr \ll \sqrt{1-r^2}$. This allows to calculate $\langle \hat{J}_z\rangle (r\pm dr) = \text{tr}[\hat{\rho}(r\pm dr,0)\hat{J}_z]$, from which we obtain the signal $\partial_r \langle \hat{J}_z\rangle$, the noise $\text{Var}[\hat{J}_z] (r) = \text{tr}[\hat{\rho}(r,0)\hat{J}^2_z]-(\text{tr}[\hat{\rho}(r,0)\hat{J}_z])^2$, and hence the sensitivity $\delta r$.
For $\delta \varphi$ around $\varphi=0$ at a desired $r$, we calculate $\hat{\rho}(r,0)$ and $\hat{\rho}(r,\pm d\varphi)$ obtaining similarly $\partial_{\varphi} \langle \hat{J}_y\rangle(r)$, $\text{Var}[\hat{J}_y](r)$, and $\delta \varphi$.

\emph{Superradiance realization: relaxation time.---}
The relaxation time to the steady-state of Eq. (\ref{ME}) can be estimated already at the mean-field level. Writing the dynamical equation for $\langle\hat{J}_-\rangle$ deduced from Eq. (\ref{ME}) and taking the collective mean-field assumption
$\langle\hat{J}_z\hat{J}_-\rangle\approx \langle\hat{J}_z \rangle \langle\hat{J}_-\rangle$, we obtain (taking $\Delta=0$) \cite{CAR,DRUMMOND1978160}
\begin{eqnarray}
\partial_t \langle\hat{J}_-\rangle \approx \gamma \langle\hat{J}_z\rangle \langle\hat{J}_-\rangle +i2\Omega \langle\hat{J}_z\rangle,
\label{MF}
\end{eqnarray}
with a corresponding equation for $\langle\hat{J}_z\rangle$. Intuitively, we can interpret the first term as an exponential relaxation of $\langle\hat{J}_-\rangle$ at a rate $\Lambda \sim \gamma (-\langle\hat{J}_z\rangle)$, whose scale we can evaluate by taking the steady-state value $-\langle\hat{J}_z\rangle \approx (N/2)\sqrt{1-r^2}$. This conclusion is supported by a thorough analysis of both the mean-field and the exact quantum dynamics \cite{Qbist}. At criticality $r=r_N$, this gives $\Lambda\sim \gamma N^{2/3}$ as quoted in the main text.

\begin{figure}[!tb]
  \centering
  \includegraphics[width=\columnwidth]{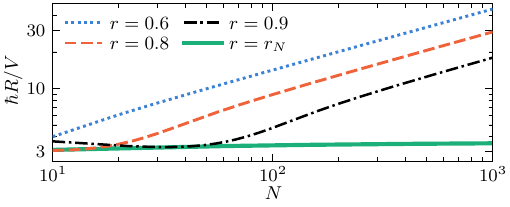}
  \caption{Adiabatic preparation of the Ising ground state [Hamiltonian (\ref{H})] by temporal sweeping of the normalized transverse field strength $r$ (taking $\beta=1$, $\varphi=0$, and with Ising coupling $V$). The upper limit $R$ on the sweep rate $\dot{r}$ is plotted against the spin number $N$ for several $r$ values. For $r<r_N$, $R$ increases with $N$, while at criticality $r=r_N$, it is roughly independent of $N$, showing no critical slowing down.
  }\label{fig5}
\end{figure}

\emph{Superradiance realization: cavity-length metrology.---}
Figure 3 establishes the equivalence between the estimation of the dipole-dipole coupling $\Delta$ and that of the CRSS phase $\varphi$, as per $\varphi=2\Delta/\gamma$, or, using the cavity parameters, $\varphi=2\delta_c/\kappa$. Here, $\delta_c=\omega_c-\omega_a$ where $\omega_a$ and $\omega_c=n (\pi/L) c$ are, respectively, the atomic and cavity resonant frequencies, while $\kappa=(\pi/L) c/\mathcal{F}$ is the cavity linewidth, with $L$ denoting the cavity length, $\mathcal{F}$ its finesse, $c$ the speed of light, and $n$ an integer. Putting these together in $\varphi=2\delta_c/\kappa$, we obtain $|d\varphi/dL| \approx 4\mathcal{F}/\lambda$, where we used $\omega_a\approx \omega_c = c 2\pi/\lambda$ after differentiation, and $\lambda$ denotes the resonant wavelength. The estimation error of $L$ is then $\delta L = \delta\varphi/|d\varphi/dL|$, obtaining the relation $\delta L/\lambda\sim \delta\varphi/\mathcal{F}$ quoted in the main text.

\emph{Ising realization: adiabatic preparation.---}
The opening, rather than closing, of the energy gap already makes a suggestive evidence of the absence of critical slowing down for adiabatic preparation. We now give a fuller treatment that supports this conclusion. Consider temporal sweeping of the transverse field strength $r(t)$ from $r(0)=0$ to $r(T)=r_N$, with the initial state being the ground state for $r=0$, $|j,-j\rangle$. The last term of the Hamiltonian $(\ref{H})$, given by $r(t)N V \hat{J}_x$ (fixing $\varphi=0$), is then time-dependent through $r(t)$. This yields the condition for adiabatic following of the ground state as, $\dot{r}< R\equiv \min_{q} (E_q- E_0)^2/(\hbar v_{q0})$ with $v_{q0}=N V |\langle q|\hat{J}_x|0\rangle|$, where $0$ and $q=1,..,N$ denote the instantaneous ground and excited states of $\hat{H}$ with energies $E_0$ and $E_q$ , respectively. To lowest-order, one only retains in $R$ the first excited state, which defines the energy gap $\varepsilon=E_1-E_0$ shown in Fig. 4d. Here, we perform numerical calculations of $R$ up to $q=3$.

In Fig. 5 we plot the upper bound $R$ on the sweep-rate $\dot{r}$ as a function of $N$ at representative $r$ values along the sweep from $r=0$ to $r=r_N$. For large enough $N$, we observe that: (i) $R$ decreases with $r$ (verified for various examples); (ii) for $r<r_N$ away from criticality, $R$ increases with $N$ roughly as $R\sim \sqrt{N}$, while at $r\sim r_N$ $R$ is effectively independent of $N$. These observations reveal that the bottleneck for fast sweeping is indeed at criticality ($r=r_N$), however, even there $R$ does not diminish with $N$. Moreover, considering the total sweep time $T\sim \int_0^{r_N} dr/\dot{r}$, whose lower bound is set by taking $\dot{r}\sim R(r)$, we can anticipate that this integral is dominated by contributions from $r<r_N$ where $R\sim \sqrt{N}$, such that $T\sim 1/\sqrt{N}$: this entails \emph{faster} rather than slower preparation time with $N$ (in analogy to the dissipative case where $T \sim N^{-2/3}$). These results agree with those we obtain analytically from a CRSS-based theory \cite{shaked}. The conclusion is that the duration of the critical metrology protocol is not limited by critical slowing down, and may be limited eventually by $N$-independent processes such as initialization.


\begin{thebibliography}{10}

\bibitem{QsenseREV}
Christian~L Degen, Friedemann Reinhard, and Paola Cappellaro.
\newblock Quantum sensing.
\newblock {\em Reviews of modern physics}, 89(3):035002, 2017.

\bibitem{QmetrologyREV}
Kok~Chuan Tan and Hyunseok Jeong.
\newblock Nonclassical light and metrological power: An introductory review.
\newblock {\em AVS Quantum Science}, 1(1), 2019.

\bibitem{OberthalerREV}
Luca Pezze, Augusto Smerzi, Markus~K Oberthaler, Roman Schmied, and Philipp
  Treutlein.
\newblock Quantum metrology with nonclassical states of atomic ensembles.
\newblock {\em Reviews of Modern Physics}, 90(3):035005, 2018.

\bibitem{YEr}
Andrew~D Ludlow, Martin~M Boyd, Jun Ye, Ekkehard Peik, and Piet~O Schmidt.
\newblock Optical atomic clocks.
\newblock {\em Reviews of Modern Physics}, 87(2):637, 2015.

\bibitem{ParisSalvatori2014}
Giulio Salvatori, Antonio Mandarino, and Matteo~GA Paris.
\newblock Quantum metrology in lipkin-meshkov-glick critical systems.
\newblock {\em Physical Review A}, 90(2):022111, 2014.

\bibitem{zanardi2008quantum}
Paolo Zanardi, Matteo~GA Paris, and Lorenzo Campos~Venuti.
\newblock Quantum criticality as a resource for quantum estimation.
\newblock {\em Physical Review A—Atomic, Molecular, and Optical Physics},
  78(4):042105, 2008.

\bibitem{Garbe2020}
Louis Garbe, Matteo Bina, Arne Keller, Matteo G.~A. Paris, and Simone
  Felicetti.
\newblock Critical quantum metrology with a finite-component quantum phase
  transition.
\newblock {\em Physical review letters}, 124(12):120504, 2020.

\bibitem{Porras2017}
Samuel Fernández-Lorenzo and Diego Porras.
\newblock Quantum sensing close to a dissipative phase transition: Symmetry
  breaking and criticality as metrological resources.
\newblock {\em Physical Review A}, 96(1):013817, 2017.

\bibitem{IvanovPorras2023}
Venelin~P. Pavlov, Diego Porras, and Peter~A. Ivanov.
\newblock Quantum metrology with critical driven-dissipative collective spin
  system.
\newblock {\em Physica Scripta}, 98(9):095103, 2023.

\bibitem{Montenegro2023}
Victor Montenegro, Marco~G. Genoni, Abolfazl Bayat, and Matteo G.~A. Paris.
\newblock Quantum metrology with boundary time crystals.
\newblock {\em Communications Physics}, 6(1):304, 2023.

\bibitem{chu2021dynamic}
Yaoming Chu, Shaoliang Zhang, Baiyi Yu, and Jianming Cai.
\newblock Dynamic framework for criticality-enhanced quantum sensing.
\newblock {\em Physical Review Letters}, 126(1):010502, 2021.

\bibitem{Marzolino2014}
Ugo Marzolino and Toma{\v{z}} Prosen.
\newblock Quantum metrology with nonequilibrium steady states of quantum spin
  chains.
\newblock {\em Physical Review A}, 90(6):062130, 2014.

\bibitem{Heugel2019}
Toni~L. Heugel, Matteo Biondi, Oded Zilberberg, and Ramasubramanian Chitra.
\newblock Quantum transducer using a parametric driven-dissipative phase
  transition.
\newblock {\em Physical review letters}, 123(17):173601, 2019.

\bibitem{ZubinQCM1}
Li-Ping Yang and Zubin Jacob.
\newblock Engineering first-order quantum phase transitions for weak signal
  detection.
\newblock {\em Journal of Applied Physics}, 126(17), 2019.

\bibitem{LesanovskyCQM2016}
Katarzyna Macieszczak, M{\u{a}}d{\u{a}}lin Gu{\c{t}}{\u{a}}, Igor Lesanovsky,
  and Juan~P Garrahan.
\newblock Dynamical phase transitions as a resource for quantum enhanced
  metrology.
\newblock {\em Physical Review A}, 93(2):022103, 2016.

\bibitem{LesanovskyCQM2024}
Albert Cabot, Federico Carollo, and Igor Lesanovsky.
\newblock Continuous sensing and parameter estimation with the boundary time
  crystal.
\newblock {\em Physical Review Letters}, 132(5):050801, 2024.

\bibitem{Plenio2022}
Theodoros Ilias, Dayou Yang, Susana~F. Huelga, and Martin~B. Plenio.
\newblock Criticality-enhanced quantum sensing via continuous measurement.
\newblock {\em PRX Quantum}, 3(1):010354, 2022.

\bibitem{Adams2022}
Dong-Sheng Ding, Zong-Kai Liu, Bao-Sen Shi, Guang-Can Guo, Klaus Molmer, and
  Charles~S. Adams.
\newblock Enhanced metrology at the critical point of a many-body rydberg
  atomic system.
\newblock {\em Nature Physics}, 18(12):1447--1452, 2022.

\bibitem{Weimer2018}
Meghana Raghunandan, J{\"o}rg Wrachtrup, and Hendrik Weimer.
\newblock High-density quantum sensing with dissipative first order
  transitions.
\newblock {\em Physical Review Letters}, 120(15):150501, 2018.

\bibitem{Clerk2022}
Martin Koppenhöfer, Peter Groszkowski, Hoi-Kwan Lau, and Aashish~A. Clerk.
\newblock Dissipative superradiant spin amplifier for enhanced quantum sensing.
\newblock {\em PRX Quantum}, 3(3):030330, 2022.

\bibitem{ClerkBelliardo2026}
Federico Belliardo, Anjun Chu, Martin Koppenh{\"o}fer, and Aashish~A Clerk.
\newblock Extracting information from a superradiant burst using simple
  measurements.
\newblock {\em PRX Quantum}, 7(4):043029, 2026.

\bibitem{Hotter2024}
Christoph Hotter, Helmut Ritsch, and Karol Gietka.
\newblock Combining critical and quantum metrology.
\newblock {\em Physical Review Letters}, 132(6):060801, 2024.

\bibitem{ivanov2020}
Peter~A Ivanov.
\newblock Enhanced two-parameter phase-space-displacement estimation close to a
  dissipative phase transition.
\newblock {\em Physical Review A}, 102(5):052611, 2020.

\bibitem{DiFresco2022}
Giovanni Di~Fresco, Bernardo Spagnolo, Davide Valenti, and Angelo Carollo.
\newblock Multiparameter quantum critical metrology.
\newblock {\em SciPost Physics}, 13(4):077, 2022.

\bibitem{MontenegroREV}
Victor Montenegro, Chiranjib Mukhopadhyay, Rozhin Yousefjani, Saubhik Sarkar,
  Utkarsh Mishra, Matteo G.~A. Paris, and Abolfazl Bayat.
\newblock Quantum metrology and sensing with many-body systems.
\newblock {\em Physics Reports}, 1134:1--62, 2025.

\bibitem{GietkaREV}
George Mihailescu, Uesli Alushi, Roberto Di~Candia, Simone Felicetti, and Karol
  Gietka.
\newblock Critical quantum sensing: A tutorial on parameter estimation near
  quantum phase transitions.
\newblock {\em PRX Quantum}, 7(2):020201, 2026.

\bibitem{Gammelmark2011}
Soren Gammelmark and Klaus Molmer.
\newblock Phase transitions and heisenberg limited metrology in an ising chain
  interacting with a single-mode cavity field.
\newblock {\em New Journal of Physics}, 13(5):053035, 2011.

\bibitem{yoav}
Yoav Shimshi and Ephraim Shahmoon.
\newblock Critical quantum metrology beyond adiabaticity in collectively pumped
  superradiance.
\newblock {\em Physical Review Letters}, 137(10):103601, 2026.

\bibitem{Frerot2018}
Irénée Frérot and Tommaso Roscilde.
\newblock Quantum critical metrology.
\newblock {\em Physical review letters}, 121(2):020402, 2018.

\bibitem{Fazio2002}
Andreas Osterloh, Luigi Amico, Giuseppe Falci, and Rosario Fazio.
\newblock Scaling of entanglement close to a quantum phase transition.
\newblock {\em Nature}, 416(6881):608--610, 2002.

\bibitem{Alejandro}
Alejandro Gonz\'alez-Tudela and Diego Porras.
\newblock Mesoscopic entanglement induced by spontaneous emission in
  solid-state quantum optics.
\newblock {\em Phys. Rev. Lett.}, 110:080502, Feb 2013.

\bibitem{Vidal2004LMG}
Julien Vidal, Guillaume Palacios, and R{\'e}my Mosseri.
\newblock Entanglement in a second-order quantum phase transition.
\newblock {\em Physical Review A}, 69(2):022107, 2004.

\bibitem{CRSS}
Ori Somech and Ephraim Shahmoon.
\newblock Quantum entangled states of a classically radiating macroscopic spin.
\newblock {\em PRX Quantum}, 5(1):010349, 2024.

\bibitem{CAR}
HJ~Carmichael.
\newblock Analytical and numerical results for the steady state in cooperative
  resonance fluorescence.
\newblock {\em Journal of Physics B: Atomic and Molecular Physics (1968-1987)},
  13(18):3551, 1980.

\bibitem{HL}
Ori Somech, Yoav Shimshi, and Ephraim Shahmoon.
\newblock Heisenberg-langevin approach to driven superradiance.
\newblock {\em Phys. Rev. A}, 108:023725, Aug 2023.

\bibitem{TOMpt}
Eric~Yilun Song, Diego Barberena, Dylan~J Young, Edwin Chaparro, Anjun Chu,
  Sanaa Agarwal, Zhijing Niu, Jeremy~T Young, Ana~Maria Rey, and James~K
  Thompson.
\newblock A dissipation-induced superradiant transition in a strontium
  cavity-qed system.
\newblock {\em Science Advances}, 11(17):eadu5799, 2025.

\bibitem{DRUMMOND1978160}
P.D. Drummond and H.J. Carmichael.
\newblock Volterra cycles and the cooperative fluorescence critical point.
\newblock {\em Optics Communications}, 27(1):160--164, 1978.

\bibitem{DRUMMOND1980}
P.~D. Drummond.
\newblock Observables and moments of cooperative resonance fluorescence.
\newblock {\em Phys. Rev. A}, 22:1179--1184, Sep 1980.

\bibitem{LAW}
RR~Puri and SV~Lawande.
\newblock Exact steady-state density operator for a collective atomic system in
  an external field.
\newblock {\em Physics Letters A}, 72(3):200--202, 1979.

\bibitem{LAR}
Julia Hannukainen and Jonas Larson.
\newblock Dissipation-driven quantum phase transitions and symmetry breaking.
\newblock {\em Physical Review A}, 98(4):042113, 2018.

\bibitem{BAR}
Diego Barberena, Robert~J Lewis-Swan, James~K Thompson, and Ana~Maria Rey.
\newblock Driven-dissipative quantum dynamics in ultra-long-lived dipoles in an
  optical cavity.
\newblock {\em Physical Review A}, 99(5):053411, 2019.

\bibitem{Lee2014}
Tony~E. Lee, Ching-Kit Chan, and Susanne~F. Yelin.
\newblock Dissipative phase transitions: Independent versus collective decay
  and spin squeezing.
\newblock {\em Physical Review A}, 90(5):052109, 2014.

\bibitem{KES}
Eric~M Kessler, Geza Giedke, Atac Imamoglu, Susanne~F Yelin, Mikhail~D Lukin,
  and J~Ignacio Cirac.
\newblock Dissipative phase transition in a central spin system.
\newblock {\em Physical Review A}, 86(1):012116, 2012.

\bibitem{RABwigner}
Julian Huber, Peter Kirton, and Peter Rabl.
\newblock Phase-space methods for simulating the dissipative many-body dynamics
  of collective spin systems.
\newblock {\em SciPost Physics}, 10(2):045, 2021.

\bibitem{MONrev}
Christopher Monroe, Wes~C Campbell, L-M Duan, Z-X Gong, Alexey~V Gorshkov,
  Paul~W Hess, Rajibul Islam, Kihwan Kim, Norbert~M Linke, Guido Pagano, et~al.
\newblock Programmable quantum simulations of spin systems with trapped ions.
\newblock {\em Reviews of Modern Physics}, 93(2):025001, 2021.

\bibitem{BlattRoos}
Rainer Blatt and Christian~F Roos.
\newblock Quantum simulations with trapped ions.
\newblock {\em Nature Physics}, 8(4):277--284, 2012.

\bibitem{Bollinger}
Joseph~W Britton, Brian~C Sawyer, Adam~C Keith, C-C~Joseph Wang, James~K
  Freericks, Hermann Uys, Michael~J Biercuk, and John~J Bollinger.
\newblock Engineered two-dimensional ising interactions in a trapped-ion
  quantum simulator with hundreds of spins.
\newblock {\em Nature}, 484(7395):489--492, 2012.

\bibitem{Ozeri}
David Schwerdt, Lee Peleg, Yotam Shapira, Nadav Priel, Yanay Florshaim, Avram
  Gross, Ayelet Zalic, Gadi Afek, Nitzan Akerman, Ady Stern, et~al.
\newblock Scalable architecture for trapped-ion quantum computing using rf
  traps and dynamic optical potentials.
\newblock {\em Physical Review X}, 14(4):041017, 2024.

\bibitem{POHrd}
N~Henkel, R~Nath, and T~Pohl.
\newblock Three-dimensional roton excitations and supersolid formation in
  rydberg-excited bose-einstein condensates.
\newblock {\em Physical review letters}, 104(19):195302, 2010.

\bibitem{Schleier_rb_2023}
Jacob~A Hines, Shankari~V Rajagopal, Gabriel~L Moreau, Michael~D Wahrman,
  Neomi~A Lewis, Ognjen Markovi{\'c}, and Monika Schleier-Smith.
\newblock Spin squeezing by rydberg dressing in an array of atomic ensembles.
\newblock {\em Physical Review Letters}, 131(6):063401, 2023.

\bibitem{Zeiher_rd_2025}
Pascal Weckesser, Kritsana Srakaew, Tizian Blatz, David Wei, Daniel Adler,
  Suchita Agrawal, Annabelle Bohrdt, Immanuel Bloch, and Johannes Zeiher.
\newblock Realization of a rydberg-dressed extended bose-hubbard model.
\newblock {\em Science}, 390(6775):849--853, 2025.

\bibitem{Kaufman_rd_2025}
Alec Cao, Theodor~Lukin Yelin, William~J Eckner, Nelson~Darkwah Oppong, and
  Adam~M Kaufman.
\newblock Autoionization-enhanced rydberg dressing by fast contaminant removal.
\newblock {\em Physical Review Letters}, 134(13):133201, 2025.

\bibitem{Heiss2005LMGsemi}
F~Leyvraz and WD~Heiss.
\newblock Large-n scaling behavior of the lipkin-meshkov-glick model.
\newblock {\em Physical review letters}, 95(5):050402, 2005.

\bibitem{Vidal2005LMG}
S{\'e}bastien Dusuel and Julien Vidal.
\newblock Continuous unitary transformations and finite-size scaling exponents
  in the lipkin-meshkov-glick model.
\newblock {\em Physical Review B—Condensed Matter and Materials Physics},
  71(22):224420, 2005.

\bibitem{Liberti2010LMG}
Giuseppe Liberti, Franco Piperno, and Francesco Plastina.
\newblock Finite-size behavior of quantum collective spin systems.
\newblock {\em Physical Review A—Atomic, Molecular, and Optical Physics},
  81(1):013818, 2010.

\bibitem{Ferioli2023}
Giovanni Ferioli, Antoine Glicenstein, Igor Ferrier-Barbut, and Antoine
  Browaeys.
\newblock A non-equilibrium superradiant phase transition in free space.
\newblock {\em Nature Physics}, pages 1--5, 2023.

\bibitem{TOMs1}
Matthew~A Norcia, Matthew~N Winchester, Julia~RK Cline, and James~K Thompson.
\newblock Superradiance on the millihertz linewidth strontium clock transition.
\newblock {\em Science advances}, 2(10):e1601231, 2016.

\bibitem{TOMs2}
Matthew~A Norcia, Julia~RK Cline, Juan~A Muniz, John~M Robinson, Ross~B Hutson,
  Akihisa Goban, G~Edward Marti, Jun Ye, and James~K Thompson.
\newblock Frequency measurements of superradiance from the strontium clock
  transition.
\newblock {\em Physical Review X}, 8(2):021036, 2018.

\bibitem{rivero_quantum_2023}
D~Rivero, C~A Pessoa~Jr, G~H De~Fran{\c c}a, R~C Teixeira, S~Slama, and Ph~W
  Courteille.
\newblock Quantum resonant optical bistability with a narrow atomic transition:
  Bistability phase diagram in the bad cavity regime.
\newblock {\em New Journal of Physics}, 25(9):093053, September 2023.

\bibitem{SKtw1}
Zhenjie Yan, Jacquelyn Ho, Yue-Hui Lu, Stuart~J Masson, Ana Asenjo-Garcia, and
  Dan~M Stamper-Kurn.
\newblock Superradiant and subradiant cavity scattering by atom arrays.
\newblock {\em Physical Review Letters}, 131(25):253603, 2023.

\bibitem{SKtw2}
Jacquelyn Ho, Yue-Hui Lu, Tai Xiang, Cosimo~C Rusconi, Stuart~J Masson, Ana
  Asenjo-Garcia, Zhenjie Yan, and Dan~M Stamper-Kurn.
\newblock Optomechanical self-organization in a mesoscopic atom array.
\newblock {\em Nature Physics}, pages 1--7, 2025.

\bibitem{MAJ}
Andreas Angerer, Kirill Streltsov, Thomas Astner, Stefan Putz, Hitoshi Sumiya,
  Shinobu Onoda, Junichi Isoya, William~J Munro, Kae Nemoto, J{\"o}rg
  Schmiedmayer, et~al.
\newblock Superradiant emission from colour centres in diamond.
\newblock {\em Nature Physics}, 14(12):1168--1172, 2018.

\bibitem{lu2026critical}
Jia-Hao L{\"u}, Wen Ning, Fan Wu, Ri-Hua Zheng, Ken Chen, Xin Zhu, Zhen-Biao
  Yang, Huai-Zhi Wu, and Shi-Biao Zheng.
\newblock Critical quantum metrology robust against dissipation and
  nonadiabaticity.
\newblock {\em Science Advances}, 12(6):eady2358, 2026.

\bibitem{Rams2018}
Marek~M. Rams, Piotr Sierant, Omyoti Dutta, Paweł Horodecki, and Jakub
  Zakrzewski.
\newblock At the limits of criticality-based quantum metrology: apparent
  super-heisenberg scaling revisited.
\newblock {\em Physical Review X}, 8(2):021022, 2018.

\bibitem{Gietka2021}
Karol Gietka, Friederike Metz, Tim Keller, and Jing Li.
\newblock Adiabatic critical quantum metrology cannot reach the heisenberg
  limit even when shortcuts to adiabaticity are applied.
\newblock {\em Quantum}, 5:489, 2021.

\bibitem{shaked}
Shaked Levy and Ephraim Shahmoon.
\newblock Finite-size theory of ising model with global interactions:
  coherently radiating spin states.
\newblock {\em manuscript in preparation}.

\bibitem{Qbist}
Nikita Leppenen and Ephraim Shahmoon.
\newblock Quantum correlated steady states under competing collective and
  individual decay.
\newblock {\em Physical Review A}, 114(1):013720, 2026.

\end{thebibliography}
\end{document}